\documentclass{elsart}
\usepackage{color}
\usepackage{graphicx}
\usepackage{amssymb}
\begin{document}
\begin{frontmatter}



\title{Electron Mobility Maximum in Dense Argon Gas at Low Temperature}


\author{A.F.Borghesani}
\address{Istituto Nazionale per la Fisica della Materia\\
Department of Physics, University of Padua\\
Via F. Marzolo8, I-35131 Padua, 
Italy\thanksref{email}}\thanks[email]{E-mail:borghesani@padova.infm.it}

\begin{abstract}
    We report measurements of excess electron mobility in dense Argon gas 
    at the two temperatures $T=152.15$ and $162.30$ K, fairly close to the 
    critical one ($T_c =150.7$~K), as a function of the gas 
    density $N$ up to $14$ atoms$\cdot$nm$^{-3}$ 
    ($N_{c}=8.08$ atoms$\cdot$nm$^{-3}$). For the first 
    time a maximum of the zero-field 
    density-normalized mobility $\mu_{0}N$ has been 
    observed at the same density where it was detected in liquid Argon
    under saturated vapor pressure conditions. The existence of the $\mu_{0}N$ maximum 
    in the liquid is commonly attributed to electrons scattering off 
    long--wavelength collective modes of the fluid, while for the 
    low--density gas a density--modified kinetic model is valid. The 
    presence of the $\mu_{0}N$ maximum also in the gas phase raises 
    therefore the question whether the single scattering picture 
    valid in the gas is valid even at liquid densities.
\end{abstract}

\end{frontmatter}
\overfullrule=20pt
\section{Introduction}
\label{Intro}
The study of the transport properties of excess electrons in dense non polar gases
gives important pieces of information on the nature, dynamics, and 
energetics of the states of excess electrons in disordered systems. 

In the 
neighborhood of the critical point of the liquid--vapor transition 
the density can be varied in a large interval with a reasonably small 
change of pressure and therefore it is relatively easy 
to investigate how the nature of the electronic states evolve starting 
from the dilute gas regime, where the kinetic theory is appropriate, towards 
the liquid regime. 

The central quantity of interest is the electron mobility $\mu.$ It 
is defined as the ratio of the mean velocity $v_{D},$ acquired 
by an electron drifting in the medium under the action of an 
externally applied and uniform 
electric field, and the strength of the field $E:\> \mu =v_{D}/E.$ Its zero--field 
limit,  $\mu_{0}= 
\lim\limits_{E\rightarrow 0}\mu ,$ is related to the fundamental 
properties of the $e-$atom interaction. In the low density regime, 
the kinetic theory relates $\mu_{0}$ to the electron--atom momentum 
transfer scattering cross section $\sigma_{mt}$ according to the 
following equation \cite{hux}
\begin{equation}
    \mu_{0}N = {4e \over 3\left(2\pi m\right)^{1/2} \left( 
    k_{\mathrm{B}}T\right)^{5/2}}\int\limits_{0}^{\infty} {\epsilon 
   \over 
  \sigma_{mt}\left(\epsilon\right) }
  \mathrm{e}^{-\epsilon/k_{\mathrm{B}}T}\, 
   \mathrm{d}\epsilon
    \label{eq:mu0kin}
\end{equation}
where $N$ is the number density of the gas. $e$ and $m$ are the 
electron charge and mass, respectively.  $T$ is the gas temperature 
and $\epsilon$ the 
electron energy.
$\mu_{0}N$ is the zero--field density--normalized mobility. 

For a given electron--atom cross section, Eq. (\ref{eq:mu0kin}) 
predicts that $\mu_{0}N$ depends only on the gas temperature and is 
independent of the gas density. 
In contrast with the prediction of the classical kinetic theory 
\cite{hux}, the electron mobility shows a strong dependence on the 
density of the medium \cite{borgpatr}. 
It is therefore important to study the effect 
of the environment on the electron--atom scattering mechanisms 
in a relatively dense phase.

Large deviations from the classical prediction, called {\sl anomalous 
density effects}, have been observed even in the simplest systems 
represented by the noble gases \cite{borgpatr}. In gases, such as He 
\cite{levine} 
and Ne \cite{borg90,borg88}, where the short--range repulsive exchange 
forces dominate the 
electron--atom interaction, there is a {\sl negative density effect}, 
namely, $\mu_{0}N$ decreases with increasing $N$ and eventually drops
rapidly to very low values because of the formation of localized 
electron states at high enough densities and in the liquid 
\cite{sakai,ssk}. 

The situation in Ar is different. Here, the $e-$atom scattering at 
low energies is essentially determined by the long--range 
attractive polarization interaction because the atomic polarizability 
of Ar is quite large. Owing to this feature, $\mu_{0}N$ shows a 
{\sl positive density effect}, i.e., it increases with increasing $N$ 
\cite{bartels}. 
A further relevant feature of Ar and of 
the heavier noble gases is that in the liquid the electron mobility 
has a value comparable to that in the crystalline state \cite{jan}. 
It is commonly believed that this characteristics of the mobility is 
due to the existence, also in the liquid, of a conduction band. 
Therefore, it is interesting to investigate, as a function of the 
density of the medium, the transition from the classical single 
scattering situation in the dilute gas to the multiple scattering 
scenario at higher densities and the eventual formation of extended 
(or localized) electron states in the liquid. 

The explanation of the different {\sl density effects} in the mobility 
is commonly based on the realization that the average interatomic 
distance at high densities becomes comparable to the electron de 
Broglie wavelength $\lambda.$ In this situation, the conditions for single 
scattering break down and quantum effects become important. Moreover, 
also the mean free path $\ell$ becomes comparable to $\lambda$ and 
multiple scattering effects come into play, too \cite{borgpatr}.

Recent and accurate measurements of mobility in Ne 
\cite{borg90,borg88} and Ar \cite{bsl,lamp94} have put into evidence 
that the different behavior of the mobility in different gases can be 
rationalized into an unified picture, where all the multiple 
scattering effects are taken into account in a heuristic way. A 
model (henceforth known as the BSL model) has been developed that 
incorporates all features of the several models proposed to 
interpret the different density effects and merges the several 
multiple scattering effects into the single scattering picture of 
kinetic theory \cite{bsl}.

Three main multiple scattering effects have been identified and all 
of them stem from the fact that the electron mean free path becomes comparable 
to its wavelength and that the latter may also become larger than the 
average interatomic distance if the density is large enough. 
The first effect is a density--dependent quantum shift $V_{0}(N)$ of 
the ground state energy of an excess electron immersed in the medium. 
According to the SJC model \cite{sjc} $V_{0}(N),$ can be written as 
\begin{equation}
    V_{0}(N) = U_{P}(N) + E_{k}(N)
    \label{eq:v0sjc}
\end{equation}
$U_{P}$ is a potential energy contribution arising from the screened 
polarization interaction of the electron with the surrounding atoms. 
$E_{k}(N)$ is a kinetic energy term, essentially due to excluded 
volume effects because the volume accessible to electrons shrinks 
as the density is increased. Owing to its nature, $E_{k}$ is positive 
and increases with increasing $N.$ An expression for it is obtained by 
imposing on the electron ground--state wave function the conditions 
of average traslational simmetry about the equivalent Wigner--Seitz 
(WS) cell centered about each atom of the gas. $V_{0}$ may be either 
$>0$ (this is the case of He \cite{broom} and Ne \cite{bmm}) or $<0$ 
(as for Ar \cite{reininger,borgcar}), depending on the relative sizes 
of $U_{P}$ and $E_{k}.$ However, the experimental mobility results 
indicate that only the kinetic energy term $E_{k}$ has to be added to the 
true electron kinetic energy when the scattering properties (namely, 
the cross sections) have to be calculated. Differently stated, the 
bottom of the electron energy distribution function is shifted by the 
amount $E_{k}$ \cite{borg88}. 

The second multiple scattering effect is an enhancement of electron 
backscattering due to quantum self--interference of the electron 
wave function scattered off atoms located along paths which are 
connected by time--reversal simmetry \cite{asca}. This phenomenon is 
closely related to the {\sl weak localization} regime of 
the electronic conduction in disordered solids and to the Anderson 
localization transition \cite{adams}. It depends on the ratio of the 
electron wavelength to its mean free path $\lambda / \ell = 
N\sigma_{mt}
\lambda.$  For Ar, at the density of the experiments, $N\sigma_{mt}
\lambda<1.$ Therefore, a linearized treatment of this effect due to 
Atrazhev {\sl et al.} can be adopted \cite{atra}. The net result is 
that the scattering cross section is enhanced by the factor 
$(1+N\sigma_{mt}\lambda/\pi ).$

 Finally, the third multiple scattering effect is due to correlations 
 among scatterers. The electron wave packet encompasses a region 
 containing several atoms, especially at low temperature and high 
 density, and is scattered off all of them simultaneously. The total 
 scattered wave packet is obtained by summing up coherently all 
 partial scattering amplitudes contributed by each atoms. The net 
 result is that the cross section is weighted by the static structure 
 factor of the fluid which is related to the gas isothermal compressibility 
 \cite{lek}.
 
 In the density--modified kinetic model (BSL model) \cite{bsl}, the 
 density--nor\-ma\-li\-zed mobility is calculated according to the 
 classical kinetic theory equations \cite{hux} with the modifications necessary 
 to take into account the mentioned multiple scattering effects
 \begin{equation}
    \mu N = -\left({e\over 3}\right)\left( {2\over 
    m}\right)^{1/2}\int\limits_{0}^{\infty}\, {\epsilon\over 
    \sigma_{mt}^{\star}\left(\epsilon+E_{k}\right)}\, 
    {\mathrm{d}g\over\mathrm{d}\epsilon}\,\mathrm{d}\epsilon
    \label{eq:muNMKM}
\end{equation}
 $g(\epsilon)$ is the Davydov--Pidduck electron energy distribution 
function \cite{cole,wan}
\begin{equation}
    g\left(\epsilon\right) = A \exp{\left\{ -\int\limits_{0}^{\epsilon}
    \left[ k_{\mathrm{B}}T + {M\over 6mz}\left( {eE\over 
    N\sigma_{mt}^{\star}}\right)^{2}\right]^{-1}
    \, 
    \mathrm{d}z\right\} 
    }
    \label{eq:davpid}
\end{equation}
where $M$ is the Ar atomic mass. $g$ is normalized as 
$\int
_{0}^{\infty}z^{1/2}g\left(z\right) \mathrm{d}z$ $=1.$  
 
The multiple scattering effects act by dressing the cross 
section so that the effective momentum--transfer scattering cross 
section is given by \cite{bsl}
\begin{equation}
    \sigma_{mt}^{\star}\left( w\right) = \mathcal{F}\left( w\right) 
    \sigma_{mt}\left( w\right)\left[ 1+{2\hbar N \mathcal{F}\left( w\right) 
   \sigma_{mt}\left( w\right) \over \left( 2 m w\right)^{1/2}}\right]
    \label{eq:smtstar}
\end{equation}
$w=\epsilon +E_{k}(N)$ is the electron energy shifted by the kinetic 
zero--point energy contribution $E_{k}.$ The group velocity of the 
wave packet is $v=[2(w-E_{k})/m]^{1/2}$ and it only contributes to the 
energy equipartition value arising from the gas temperature 
\cite{wan}. 
The Lekner factor $\mathcal{F}$ \cite{lek} takes into account correlations 
among scatterers 
\begin{equation}
    \mathcal{F}\left( k\right) ={1\over 4k^{4}}\int\limits_{0}^{2k} 
    q^{3}S(q) \, \mathrm{d}q
    \label{eq:lekner}
\end{equation} with $k^{2}=2m\epsilon / \hbar^{2}.$
This is equivalent to the formulation given elsewhere \cite{nakamura}.
An expression for the static structure factor in the 
Orn\-stein--Zer\-nic\-ke form in the near--critical region of Ar has been 
found in literature \cite{ts}
\begin{equation}
    S\left( q\right)={S\left( 0\right) +\left( qL\right)^{2}\over 1+ 
    \left( qL\right)^{2}}
    \label{eq:sdiq}
\end{equation} where $S(0)$ is related to the gas isothermal 
compressibility $\chi_{T}$ by the relation 
 $   S\left(0\right)= Nk_{\mathrm{B}}T\chi_{T}
 $ and 
the correlation length $L$ is defined as $
    L^{2}=0.1 l^{2}\left[S\left( 0\right) -1\right]
.$ $l\approx 10\,\mbox{\AA}$ is the so--called {\sl short--range 
correlation length} \cite{ts}.

Experiments in Ne at $T\approx 45$ K \cite{borg90,borg88} and in Ar 
at $T=162.7 $ K up to $N\approx 6.5\>\, 
\mathrm{atoms}\cdot\mathrm{nm}^{-3}$ \cite{bsl} proved 
that the kinetic energy shift can be calculated according to the 
Wigner--Seitz model \cite{ws} as $E_{k}=E_{WS}\equiv 
\hbar^{2}k_{0}^{2}/2m,$ where the wavevector $k_{0}$ is obtained by 
self--consistently solving the eigenvalue equation
\begin{equation}
    \tan{\left[ k_{0}\left( r_{s} - 
    \tilde a \left( k_{0}\right)\right)\right]}
    -k_{0}r_{s}=0
    \label{eq:ews}
\end{equation}
$r_{s}=(3/4\pi N)^{1/3}$ is the radius of the WS cell and 
$\tilde a$ is the hard--core radius of the Hartree--Fock potential 
for rare gas atoms. In the BSL model, according to a suggestion found in 
literature \cite{sjc}, $\tilde a$ 
is estimated from the total 
scattering cross section as $\tilde a = \sqrt{\sigma_{T} /4\pi}.$

The BSL model has been successfully used to analyze the experimental 
data in Ar at $T=152.5 $ K up to $N\approx 10  \> \, 
\mathrm{atoms}\cdot\mathrm{nm}^{-3}$ \cite{lamp94} by assuming that 
for the highest densities the WS model for the calculation of $E_{k}$ 
is inappropriate and $E_{k}$ must be deduced from the experiment. 

It is natural to ask if the reason of the deviation of the 
experimentally determined $E_{k}$ values from the WS model
is that the BSL model has been used beyond its limits of 
applicability, or if different mechanisms become active for momentum 
transfer processes at high $N.$ It is known, in fact, that in liquid 
Ar the mobility of thermal electrons shows a maximum not very distant 
from the highest density investigated in previous experiments 
\cite{lamp94}. The mobility maximum occurs practically at the same 
density where $V_{0}(N)$ has a minimum \cite{reininger}. The existence of 
this maximum, that indicates a situation of minimum scattering, has 
been interpreted within the deformation--potential theory 
\cite{basak,nishi}. 
Intrinsic 
density fluctuations of the fluid modulate the energy $V_{0}$ at the 
bottom of the conduction band. The spatial disuniformity of the 
ground--state energy of electrons is the source of scattering. This 
is an intrinsic multiple--scattering theory because electrons are 
scattered off long--wavelength collective modes of the fluid. 
These sorts of {\sl phononic models} \cite{basak,nishi,naveh} do correctly 
predict the existence of the mobility maximum at the right value 
$N\approx 12.5 \>\, \mathrm{atoms}\cdot\mathrm{nm}^{-3},$ 
but they 
fail to predict the density- and electric field dependence of the 
$\mu N$ data as the BSL model does. 

Owing to these reasons, we have investigated an extended density range 
in Ar at $T=152.15 $ K up to a maximum density $N=14\>\, 
\mathrm{atoms}\cdot\mathrm{nm}^{-3},$ well above the density of the 
mobility maximum in liquid, in order to see if the mobility maximum is 
a feature typical of the liquid only or if it appears also in the gas. 
In the latter case, there would be sound motivations to extend the 
density-modified kinetic approach even to the liquid 
\cite{kaneko,iannuzzi,nakamura}.

\section{Experimental Details}\label{tech}
The electron mobility measurements have been carried out by using the 
well--known pulsed photoemission technique \cite{borg88,borg96}.
A swarm of electrons drifts in the gas under the action of an 
externally applied uniform electric field. The time $\tau_{e}$ spent by the 
particles crossing the drift distance $d$ is measured and the drift 
velocity is calculated as $v_{D}=d/\tau_{e}.$ The mobility is simply 
obtained as $\mu=v_{D}/E,$ where $E$ is the strength of the electric 
field. 
\begin{figure}[htbp]
    \centering
    \includegraphics[scale=0.5]{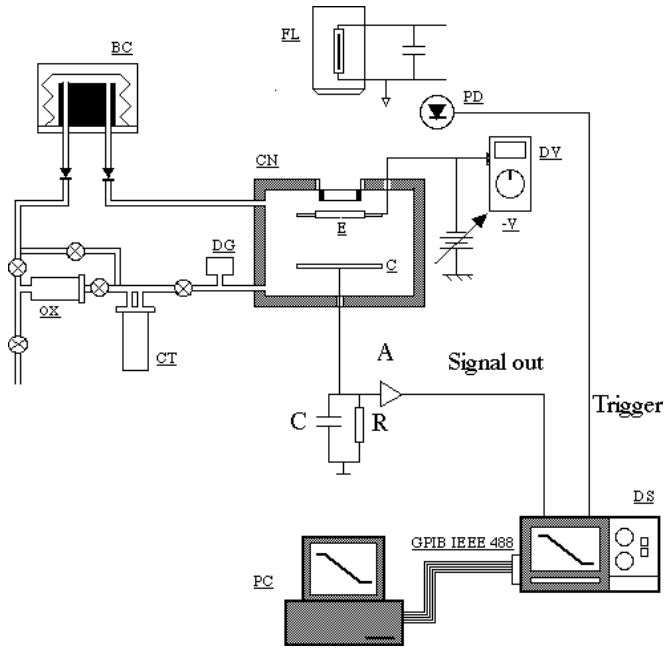}
    \caption{\small Simplified schematics of the experimental setup. 
    FL: Xe flashlamp, CN: cell, PD: photodiode, BC: bellow circulator, 
    OX: Oxisorb cartridge, DV: digital voltmeter, -V: high--voltage 
    generator, CT: LN2--cooled active charcoal trap, DG: pressure 
    gauge, PC: personal computer, E: emitter, C: collector, A: 
    amplifier, DS:digital scope.}
    \label{fig:apparatus}
\end{figure}
A simplified schematics of the experimental apparatus is shown in 
Fig. 1.
A high--pressure cell is mounted on the cold head of a cryocooler 
inside a triple--shield thermostat. The cell can be operated in the 
range $25<T<330 $ K and its temperature can be stabilized within $\pm 
0.01$ K. The cell can withstand pressures up to 10 MPa and the 
pressure $P$ is measured with an accuracy of $\pm 1$ kPa. The gas 
density $N$ is calculated from $T $ and $P$ by means of an accurate 
equation of state \cite{wag}.

A parallel--plate capacitor is contained in the cell and is powered 
by a D.C. high--voltage generator \cite{delfitto90}. A 
gold--coated fused silica window is placed in the center of one of the 
plates and can be irradiated  with a short pulse $(\approx 4\, 
\mu\mathrm{s})$ of VUV light produced by a Xe flashlamp. Thus, a 
bunch of electrons, whose number ranges between 4 and 400 fC 
depending on the gas pressure and on the applied electric field strength, 
is photoinjected into the drift space. During the drift motion of the 
charges towards the anode a current is induced in the external circuit. 
In order to improve the signal--to--noise ratio the current is 
integrated by a passive RC network. The resulting voltage 
signal is amplified and recorded by a digital oscilloscope. The drift 
time is obtained by analyzing the signal waveform with a personal 
computer. Typical signal waveforms are shown in Fig. 2. 
The estimated error on the mobility is less than $ 5\, \% .$
\begin{figure}[htbp]
    \centering
    \includegraphics[scale=0.75]{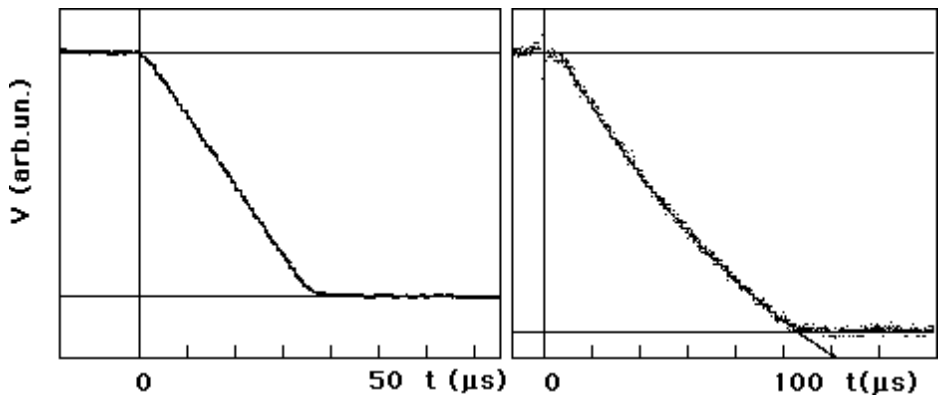}
    \caption{\small Signal waveforms induced by electrons uniformly 
    drifting in a gas. At left: pure gas, at right: gas containing a 
    few p.p.m. of O$_{2}$ electron--attaching impurities.}
    \label{fig:wf}
\end{figure}
\par\noindent The gas used is ultra--high purity Argon with a nominal 
O$_{2}$ content of 1 part per million. Further purification is 
accomplished by recirculating the gas in a closed loop through a 
LN2--cooled activated--charcoal trap and a commercial Oxisorb 
cartridge. The final O$_{2}$ amount is estimated to be a fraction of a 
part per billion. 
\section{Experimental Results and Discussion}\label{res}
We have carried out measurements at two different temperatures in the 
neighborhood of the critical temperature, namely, at $T=162.30$ K and 
$T=152.15$ K $(T_{c}=150.7\, \mathrm{K}).$ We have investigated the 
dependence of the density--normalized mobility $\mu N$ as a function 
of the density--reduced electric field strength $E/N$ and of the 
density $N.$ The density range explored encompasses the critical 
density $N_{c}= 8.08\,\, \mathrm{atoms}\cdot\mathrm{nm}^{-3}.$ 
In Fig. 3 we show sample $\mu N$ 
data at $T=162.30$~K. 
\begin{figure}[htbp]
    \centering
    \includegraphics[scale=0.45]{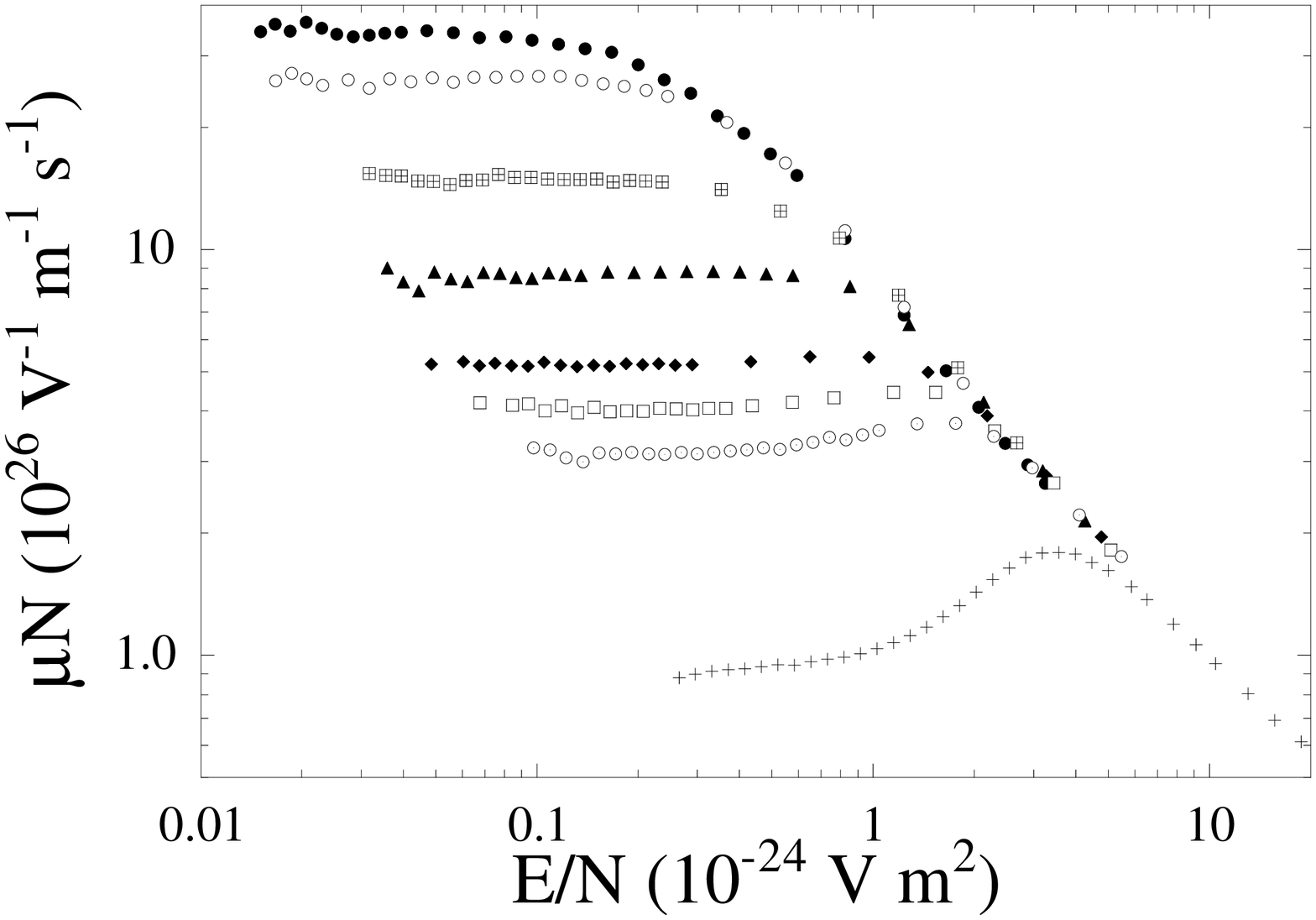}
    \caption{\small $\mu N$ as a function of the reduced electric 
    field strength $E/N$ for $T=162.30$ K. The densities are (from 
    top) $N=11.74,\, 11.15,\, 9.96,\,  9.07,\, 8.08,\, 7.58,$ $ 7.06,\, 
    5.03\,\, \mathrm{atoms}\cdot\mathrm{nm}^{-3}.$}
    \label{fig:muNdiEN162}
\end{figure}
These data agree well with previous measurements at $T=162.7$ 
K \cite{bsl}. The data for $T=152.15$ K are qualitatively similar to those 
shown in Fig. 3.

The behavior of the reduced mobility $\mu N$ of excess electrons 
in Ar as a function of the reduced electric field is quite 
complicated, although it is now well understood for not too 
high densities $(N< 7.0\> \mathrm{atoms}\cdot\mathrm{nm}^{-3})$ \cite{bsl}.
At low field strength $\mu N$ is a constant independent of 
$E/N.$ This constant value is the zero--field density--normalized 
mobility $\mu_{0}N$ and pertains to thermal electrons. In fact, at 
such low fields, the energy gained by electrons from the field is 
negligible in comparison with their thermal energy. According to the 
prediction of the classical kinetic theory \cite{hux}, $\mu_{0}N$ should be 
constant and independent of $N,$ while, experimentally, a completely 
different behavior of $\mu_{0}N$ is observed, as it is easily 
realized by observing Fig. 3.

At small and medium $N,$ $\mu N$ displays a maximum as a function of 
$E/N$ in the range $(2\leq E/N\leq 4) \times 
10^{-24}\,\mathrm{V}\,\mathrm{m}^{2},$ whose position depends on the density. 
Since the maximum is observed from the dilute gas up to the present densities, it 
is obvious that it has to be attributed to the
Ramsauer--Townsend minimum of the $e-$Ar atomic momentum--transfer scattering cross 
section, which is located at an electron energy $\epsilon_{RT}\approx 230 $ 
meV \cite{wey,had}. When $E/N$ has the value $(E/N)_{max}$ 
corresponding to the mobility maximum, the average electron energy is 
equal to the energy of the Ramsauer minimum $\langle \epsilon 
\rangle =\epsilon_{RT}.$ For larger $N,$ the mobility maximum at $(E/N)_{max}$ 
gradually disappears.

Finally, for even larger $E/N\geq 4\times 10^{-24}\, \mathrm{V}\, 
\mathrm{m}^{2},$ the $\mu N$ curves for all densities merge into a 
single curve that is well described by the classical kinetic 
equations with the given cross section. 
For large $E/N,$ the behavior of $\mu N$ becomes therefore independent of 
density and is easily explained in terms of the BSL model. At low 
$E/N$, i.e., at small electron energies $\epsilon,$ the  
de Broglie wavelength of the electron, $\lambda =h/\sqrt{2m\epsilon},$ 
is pretty large 
and the electron wavepacket is so much extended as to encompass many 
atoms at once. In this situation, multiple scattering effects are very 
important. As the mean electron energy $\langle \epsilon\rangle$ is 
increased by increasing $E/N,$ the extension of the electron 
wave function, as measured by $\lambda,$ decreases and the importance 
of multiple scattering is reduced. Therefore, the experimental points 
converge to the prediction of the classical kinetic theory. This 
behavior is present also in the liquid \cite{lamp94} and has been 
observed also in Neon gas \cite{borg88}, where the energy dependence 
of the cross section is completely different from that of Ar and where 
the temperature of the experiment $(T\approx 45\>\mathrm{K})$ was much 
lower than in the present case.

As already observed for several temperatures \cite{bsl,lamp94}, the mobility 
maximum related to the Ramsauer minimum of the cross section shifts 
to lower $E/N$ values as the density is increased, as shown in Fig. 4, where the reduced field of the mobility 
maximum, $(E/N)_{max},$ is plotted as a function of $N$ for $T=152.15 $ K.
This behavior is also consistent with previous measurements in 
gaseous Ar at room temperature for densities up to $2\>\mathrm{atoms}\cdot 
\mathrm{nm}^{-3}$ \cite{christo}. 
\begin{figure}[htbp]
    \centering
    \includegraphics[scale=0.5]{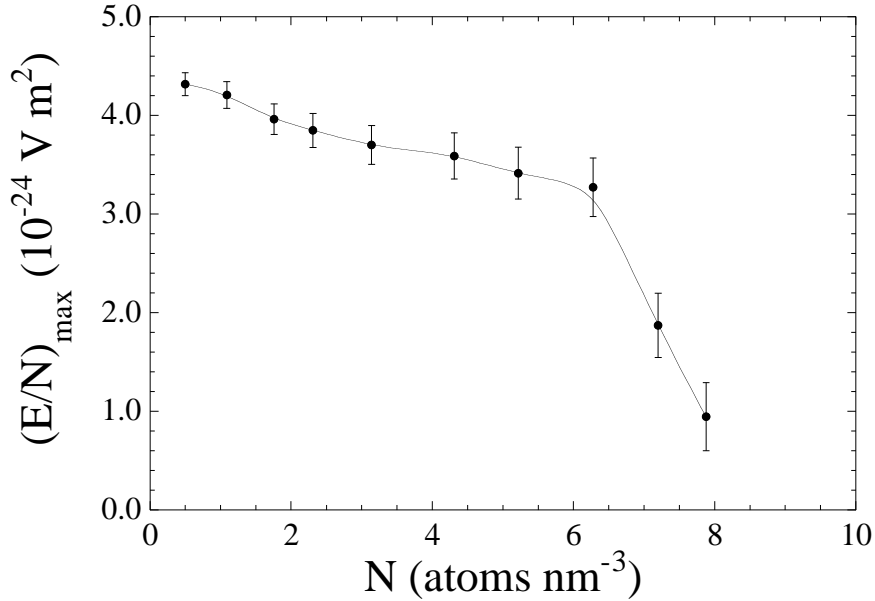}
    \caption{\small Decrease of $(E/N)_{max}$ with increasing $N$ for 
    $T=152.15 $ K. The line has no theoretical meaning.}
    \label{fig:esunmax}
\end{figure}
On approaching the critical density, for $N>6.0 \,\> 
\mathrm{atoms}\cdot\mathrm{nm}^{-3},$ the decrease of $(E/N)_{max}$ 
proceeds at a much faster rate than before, until, for $N\approx N_{c}=8.08 
\>\mathrm{atoms}\cdot \mathrm{nm}^{-3}, $ $(E/N)_{max}\rightarrow 0$ 
and the mobility maximum disappears, as can be seen in Fig. 3.

The observed behavior can be explained by noting that, for 
$(E/N)_{max},$ the average electron energy equals that of the Ramsauer 
minimum of the cross section, $\langle \epsilon\rangle  =\epsilon_{RT}.$ 
Generally speaking, it turns out that $\langle \epsilon\rangle$ can 
be written in the form
\begin{equation}
    \langle \epsilon\rangle  = \frac{3}{2} k_{\mathrm{B}}T +E_{k}(N) + 
    f\left( E/N\right)
    \label{eq:eave}
\end{equation} where $f\left( E/N\right)$ is a monotonically 
increasing function of the reduced electric field \cite{wan}. Since 
the density--dependent electron kinetic energy shift $E_{k}(N)$ 
increases with increasing $N,$ 
$(E/N)_{max}$ must decrease in order to fulfill the condition 
$\langle \epsilon\rangle = \epsilon_{RT} $ with 
increasing $N$ at constant temperature. 
[We will furthermore show that the change of slope of $(E/N)_{max}$ as 
a function of $N$ for $N>6.0 \> \mathrm{atoms}\cdot \mathrm{nm}^{-3}$ 
is related to the change of slope of $E_{k}(N)$ at the same density.] 
Eventually, for $N> N_{c},$ the electron energy distribution function 
is so largely shifted by $E_{k}$ as to sample the scattering cross 
section well beyond the Ramsauer--Townsend minimum. This is the 
reason of the disappearing of the mobility maximum. 

A cornerstone for the understanding of the electron scattering in 
dense gases is represented by the analysis of behavior of the zero--field 
density--normalized mobility $\mu_{0}N$ 
\begin{figure}[htbp]
    \centering
    \includegraphics[scale=0.45]{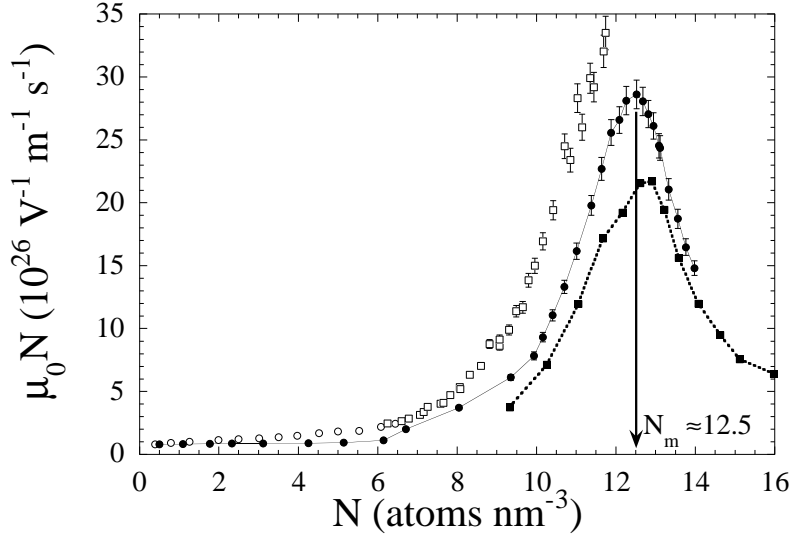}
    \caption{\small $\mu_{0}N$ as a function of $N$ for $T=162.30$ 
    (open squares), 
    $T=162.7$ (open circles) \cite{bsl},  
    and $152.15$ K (closed circles). The closed squares are the 
    results in liquid Ar \cite{lamp94}. The arrow indicates the value 
    of the density $N_{m}$ of the $\mu_{0}N$ maximum in gas. The lines 
    are only guides for the eye.}
    \label{fig:confrontoArGasLiq}
\end{figure}
\noindent as a function of the density, because, as 
already pointed out, the classical kinetic theory predicts that 
$\mu_{0}N$ is independent of $N,$ while the experiment gives a 
strongly density--dependent $\mu_{0}N$ for every explored temperature.
In Fig. 5 we therefore show the zero--field density--normalized 
mobility $\mu_{0}N$ as a function of $N$ for the two investigated 
temperatures $T=162.30$ K and $T=152.15 $ K. Previous data taken at 
$T=162.7$ K with a different apparatus \cite{bsl} are reported in order 
to show the consistency of the present data. Moreover, also data 
obtained in liquid Ar \cite{lamp94} are shown for comparison, although 
it must be remembered that the measurements in the liquid are taken along
the liquid-vapor coexistence line and are therefore not isothermal.

Two relevant features emerge from Fig. 5. The first one 
is the impressive increase of $\mu_{0}N$ with increasing $N$ for both 
temperatures up to $N\approx 
11.0\>\mathrm{atoms}\cdot\mathrm{nm}^{-3}.$ This behavior is present 
also at room temperatures \cite{bartels} and has been one the primary 
motivations for the development of multiple scattering theories.
The BSL model explains quantitatively this feature of $\mu_{0}N $ 
for $N\leq 10 \> \mathrm{atoms}\cdot\mathrm{nm}^{-3}.$ For 
$E/N\rightarrow 0,$ electrons are in thermal equilibrium with the 
host gas and do not gain energy from the field. Therefore, their 
average energy is $\langle 
\epsilon\rangle \ll \epsilon_{RT}.$ In this energy range, the 
momentum--transfer cross section decreases rapidly with increasing 
electron energy \cite{wey,had}, as shown in Fig. 6.
\begin{figure}[htbp]
    \centering
    \includegraphics[scale=0.45]{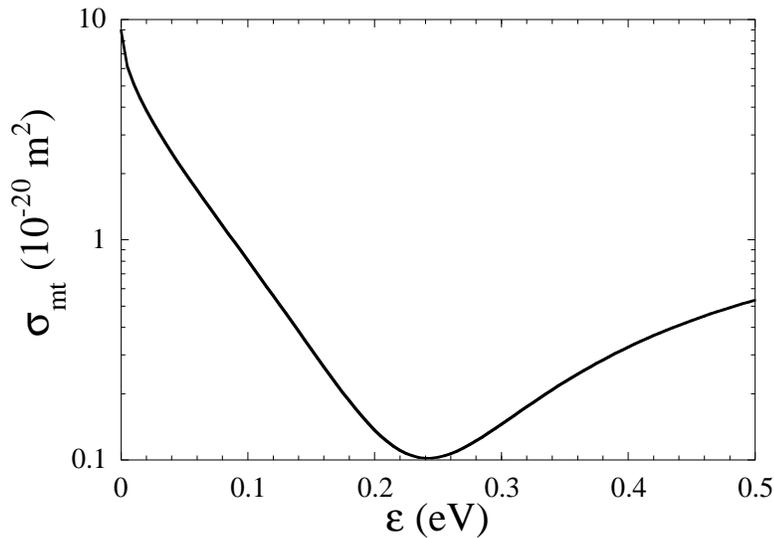}
    \caption{\small Momentum--transfer scattering cross section of Ar 
    \cite{wey}.}
    \label{fig:smt}
\end{figure}
\noindent Roughly speaking, $\mu_{0}N$ is a sort of weighted average of the 
inverse cross section, as it can be realized by inspecting Eq. 
\ref{eq:mu0kin}. To a first approximation, $\langle 1/\sigma_{mt}\rangle 
\approx 1/\sigma_{mt}\left(\langle \epsilon \rangle\right).$ At 
constant $T,$ this average should be constant and should not depend 
on the density $N,$ unless $\langle 
\epsilon \rangle$ depends on it. In particular, owing to the shape of 
$\sigma_{mt}(
\epsilon) ,$ $\mu_{0}N$ can increase with increasing $N$ only if 
$\langle \epsilon\rangle$ does the 
same. Therefore, the positive density effect of $\mu_{0}N$ supports 
the conclusion that the average electron kinetic energy includes a 
density--dependent contribution that is positive and increases with 
increasing $N,$ as expressed by Eq. \ref{eq:eave}. This conclusion 
immediately rationalizes the observations described by Christophorou 
{\sl et al.} \cite{christo}. In fact, they note that the minimum of 
the inverse density--normalized mobility at zero--field, i.e. a 
quantity proportional to the cross section, shifts to lower energies 
as the density is increased. They calculate the average electron 
energy at zero field according to the Nernst--Einstein--Townsend relation \[ 
\langle \epsilon\rangle ={3\over 2} {D_{L}\over \mu_{0}} = {3\over 
2}k_{\mathrm{B}}T\] where $D_{L}$ is the longitudinal diffusion 
coefficient. By so doing, they neglect the density--dependent 
zero--point electron energy $E_{k}(N),$ and therefore the 
Ramsauer--Townsend minimum seemingly appears at lower energies, as 
can be seen by inspecting Eq. \ref{eq:eave} with 
$E/N=0,\>\,\mathrm{i.e.} (f(E/N)=0),$ 
and by neglecting $E_{k}.$ 

The second most relevant feature is the presence of a sharp maximum 
of $\mu_{0}N$ for $N\equiv N_{m}\approx 12.5 \> 
\mathrm{atoms}\cdot\mathrm{nm}^{-3} $ at $T=152.15 $ K. [Probably, 
such maximum would exist also at the higher temperature, but the 
limited pressure range of the experimental cell $(P\leq 10.0\,\> 
\mathrm{MPa})$ did not allow the investigation of larger densities at 
higher temperatures.] The $\mu_{0}N$ maximum occurs at nearly the 
same density where it was observed in liquid Ar \cite{lamp94}, as also shown 
in Fig. 5. 
A similar behavior has been previously observed in liquid and gaseous CH$_{4}$ \cite{hol}.

It is well--known \cite{basak,naveh} that the maximum of $\mu_{0}N$ 
in liquid has been attributed to scattering of electrons off 
long-wavelength collective modes of the fluid. The intrinsic density 
fluctuations of the liquid modulate the electron energy at the bottom 
of the conduction band of the liquid, $V_{0}(N),$ and the spatial 
fluctuations of $V_{0}(N)$ act as the potential for the scattering, 
just as lattice deformation--potentials from acoustic phonons scatter 
carriers in semiconductors. 
Within the {\sl deformation potential} theory, the potential for 
scattering is linear in the density deviations about the average value. 
Since it happens that $V_{0}(N)$ has a minimum at nearly the same density, 
$N_{m},$ where $\mu_{0}N$ is maximum \cite{reininger}, the scattering 
potential vanishes to first order at $N_{m}.$ Therefore, at this 
density the scattering of electrons is very much reduced and the electron mobility 
turns out to be maximal. For any other $N\neq N_{m},$ the slope of 
$V_{0}(N)$ as a function of $N$ is nonzero and deformation potential 
is large enough to efficiently scatter electrons, thus reducing their 
mobility. 

However, a gas does not support phonons as a liquid does. Therefore, 
the presence of the $\mu_{0}N$ maximum in dense Ar gas at the same 
density $N_{m}$ as in the liquid raises the question if, beyond a given 
density threshold, there is a 
change in the physical mechanisms which determine the mobility in the 
gas, or, rather, if the single 
scattering picture of the density--modified kinetic theory can be 
extended to the liquid. 

On one hand, one could argue that, at low 
and medium densities, the single scatterer approximation is valid and 
electrons can be described as scattering off individual atoms, though 
the scattering properties have to be density--modified in order to account for 
multiple scattering effects. On increasing the density, a conduction 
band might develop and electrons might be scattered off 
long--wavelength collective modes of the dense gas, though they might 
not be true phonons. 

On the other hand, there are several reasons to extend the single 
scattering picture to the liquid. First of all, the phononic theories 
are developed for thermal electrons only, i.e., they make predictions 
only on $\mu_{0}N$ and completely disregard the electric field 
dependence of the experimental $\mu N$ data. This dependence is very 
important because it is intimately related to the shape of the atomic 
cross section. There have indeed been more or less successful attempts to use the classical 
kinetic theory even in the liquid \cite{kaneko,iannuzzi,nakamura}, 
though the cross section has been taken as an adjustable 
parameter. Moreover, the {\sl phononic} models do not even predict 
accurately $\mu_{0}N,$ unless higher--order scattering processes 
are taken into account \cite{nishi,naveh}. 

It is therefore challenging to investigate the possibility that the 
density--modified kinetic model can account for the $\mu_{0}N$ 
maximum and that a single--scattering picture can be retained even in 
the liquid, owing to its conceptual simplicity.
The BSL model has been thus used for the analysis of the experimental 
data at high density. 

First of all, we do not make any assumptions on the value of the 
kinetic energy shift $E_{k}(N).$ We treat 
it as an adjustable parameter to be determined by fitting the equations of 
the model Eqns. \ref{eq:muNMKM}--\ref{eq:smtstar} in the limit $E/N\rightarrow 0$ to the 
experimental data. Literature data for the cross section have been 
used \cite{wey}. Once $E_{k}(N)$ has been determined by this fitting 
procedure, the average electron energy (Eq. \ref{eq:eave}) could 
be evaluated, if necessary, as 
\[ \langle \epsilon\rangle =E_{k}(N)+\int\limits_{0}^{\infty}z^{3/2}g(z) 
\mathrm{d}z  \] where $g$ is given by Eq. \ref{eq:davpid}, and Eq. 
\ref{eq:eave} is recovered.
\begin{figure}[htbp]
    \centering
    \includegraphics[scale=0.45]{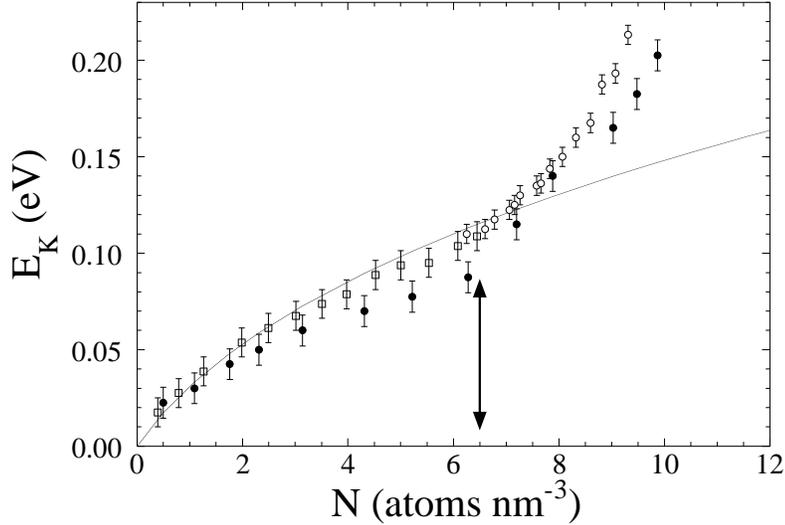}
    \caption{\small Values of $E_{k}$ plotted as a function of $N$ for 
    $T=162.30$ K (open circles) and $T=152.15 $ K (closed circles). 
    Previous values determined from 
    data at $T=162.7 $ K \cite{bsl} are shown (open squares). The 
    solid line is the prediction of the WS model.
    The arrow indicates the density where 
    $(E/N)_{max}$ changes slope.}
    \label{fig:ek}
\end{figure}
In Fig. 7 the resulting $E_{k}$ values are shown as a function of 
$N$ for the two investigated temperatures. Previous determinations of 
$E_{k}$ for $T=162.7$ K \cite{bsl} are also shown for comparison to assess the 
consistency of these new sets of measurements with previous ones.

The data at $T=162.30 $ K agree very well with the data taken at 
$T=162.7 $ K. There are small differences between the results for 
$T=152.15 $ K and $T=162.30$ K, which might be attributed to the 
larger gas compressibility for the temperature close to $T_{c}$ and to 
the uncertainty with which the {\sl short--range correlation 
length} $l$ is known \cite{ts}. Nonetheless, the experimentally 
determined $E_{k}$ values agree pretty well with the prediction of 
the WS model (shown as a solid line in Fig. 7) for 
densities up to $\approx 7.0 \> \, 
\mathrm{atoms}\cdot\mathrm{nm}^{-3}.$ For larger $N,$ starting at practically 
the same density where $(E/N)_{max}$ changes slope (see Fig. 4), up to $N\approx 10\>\, 
\mathrm{atoms}\cdot\mathrm{nm}^{-3},$ the values of $E_{k}$ that 
reproduce the experimental values of $\mu_{0}N$ increase faster 
with $N$ than the prediction of the WS model. This is not a failure of 
the BSL model. It is just related to the fact that the WS model is 
only valid when $r_{s}\gg \tilde a .$ Unfortunately, up to now there are no 
theoretical calculations of $E_{k}$ with which the present results 
can be compared. 

In the density range of the present experiment, the BSL model does not reproduce only 
$\mu_{0}N$ with the proper choice of $E_{k}(N).$ It also shows a high 
degree of internal consistency because the value of $E_{k}$ 
determined from the data at low field allows to reproduce quite 
accurately the full $E/N-$ dependence of $\mu N. $ This result is 
shown in Fig. 8,
\begin{figure}[htbp]
    \centering
    \includegraphics[scale=0.45]{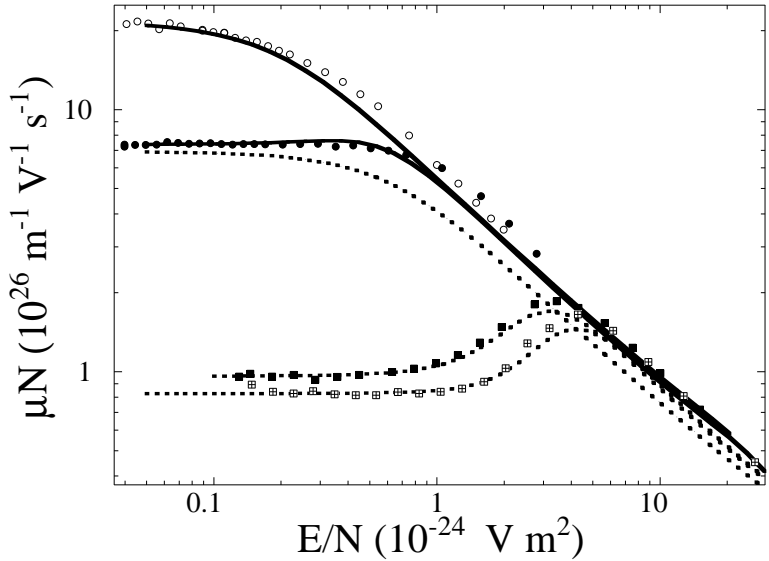}
    \caption{\small Comparison of the results of the BSL model with 
    the experimental $\mu N(E/N)$ data for some densities (dotted 
    lines). The densities are (from top): $N= 13.33,\, 
    9.935,\, 5.144,\, 0.502$ $\mathrm{atoms}\cdot\mathrm{nm}^{-3}.$ 
    The solid lines are obtained by using the effective cross 
    section $\sigma_{eff},$ as described in the text.}
    \label{fig:muncalc}
\end{figure}
where the curves for several densities are compared with the 
prediction of the BSL model (dotted lines). The dotted curves are 
obtained by using the $E_{k}(N)$ values determined by fitting the model to 
the zero--field data. 
It can be realized that, at small and medium 
$N,$ the features of the mobility are all reproduced 
well. Namely, the position and strength of the mobility maximum as well as its 
behavior as a function of the density are described accurately. The 
behavior at small-- and high--fields of $\mu N$ is correctly 
reproduced, with the curves for different densities merging into a 
single one at large $E/N.$ All these observations are consistent with 
the hypothesis that the kinetic--energy shift $E_{k}(N)$ increases 
with increasing $N,$ and that it can grow so large as to shift the 
average electron energy beyond $\epsilon_{RT}.$

However, it is also evident that, for even larger $N\, (\geq 10\>\, 
\mathrm{atoms}\cdot\mathrm{nm}^{-3}), $ the BSL model as such does 
neither reproduce the $\mu_{0}N$ maximum for $N=N_{m},$ nor the decrease of 
$\mu_{0}N$ with increasing $N$ for $N>N_{m}.$ 
On one hand, it is clear, of course, 
that the overall behavior of $\mu_{0}N$ as a function of $N$ must be 
related to the shape of the atomic cross section and to the 
density--dependent quantum shift of the electron energy distribution function. 
Unfortunately, the scattering cross sections are known with limited accuracy 
as far as strength and position of the Ramsauer minimum are concerned. 
Different choices of $\sigma_{mt}$ give different strength and 
position of the $\mu_{0}N$ maximum \cite{lamp94} or they may not even give a 
maximum at all.

On the other hand, the use of an effective density--modified scattering 
cross section $\sigma_{mt}^{\star}$ has proven so powerful giving a 
nice agreement between model and  data up to fairly large 
densities that it is interesting to extend this paradigm to higher 
densities. 
According to a suggestion proposed in literature 
\cite{lamp94,kaneko}, at high $N$ a good agreement between data and model can 
be obtained by  scaling $\sigma_{mt}^{\star}$ by a 
factor $c_{0}(N).$ The adjustable parameter $c_{0}$ depends only on $N,$ but it is 
independent of $E/N$ and of the electron energy. Therefore, it has no 
influence on the dependence of $\mu N$ on $E/N$ at constant density. 
While $c_{0}$ is introduced as an adjustable parameter, the energy 
shift $E_{k}$ is no longer determined from the experimental data. 
Instead of $E_{k}$ it is rather used the value $E_{WS}$ given by the WS 
model and calculated according to Eq. \ref{eq:ews} (solid line in 
Fig. 7). \begin{figure}[htbp]
    \centering
    \includegraphics[scale=0.4]{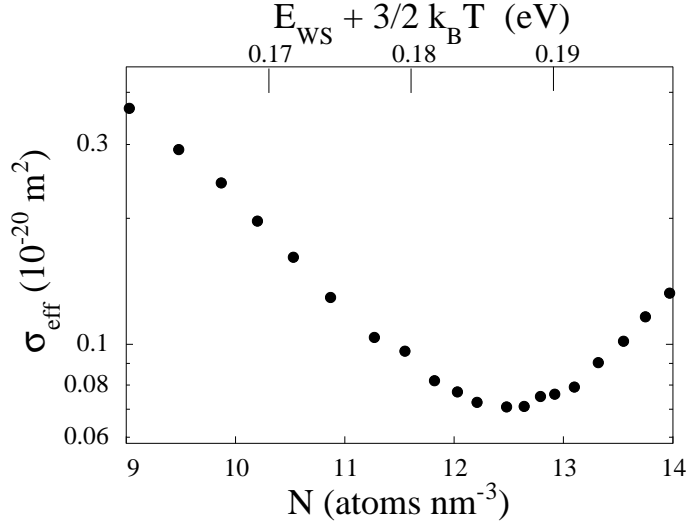}
    \caption{\small Effective scattering cross section 
    $\sigma_{eff}=c_{0}(N)\sigma_{mt}^{\star}$ as a function of $N$ 
    (lower scale) and energy (upper scale). The upper scale has been 
    obtained by converting density to energy by means of the WS model.}
    \label{fig:seff}
\end{figure}
$\sigma_{mt}^{\star}$ is everywhere replaced by 
$\sigma_{eff}=c_{0}\left( N \right)\sigma_{mt}^{\star}$ 
and $c_{0}$ is so adjusted as to reproduce the behavior of $\mu_{0} N$ 
as a function of $N.$ $c_{0}$ turns out to be of order unity 
$c_{0}\approx \mathcal{O}(1).$ With this choice the electric field 
dependence of $\mu N$ is reproduced even at the highest densities.  
The shape of the effective cross section at thermal 
energies $\sigma_{eff}=c_{0} (N) \sigma_{mt}^{\star} [ (3/2) 
k_{\mathrm{B}} T + E_{WS} (N) ] $ is shown in Fig. 9.
The energy scale on the upper horizontal axis has been obtained by 
converting density to energy by means of the WS model Eq. \ref{eq:ews}. 
By comparing Fig. 9 to Fig. 6 there is 
undoubtely a strong similarity between $\sigma_{eff}$ and the atomic 
cross section $\sigma_{mt}.$ In particular, it is interesting the 
presence of a Ramsauer--type minimum also in the effective cross section. 
Moreover, the strength of $\sigma_{eff}$ is very close to its atomic 
companion, though the position of the minimum occurs at a somewhat 
lower energy and the minimum itself appears to be narrower. This feature might be due to 
the use of the WS model for the $N\rightarrow \epsilon$ conversion. 
If the experimentally determined values of $E_{k}(N)$ had been used 
instead of the WS model, the $\sigma_{eff}$ minimum would be broader 
and shifted to larger energies because $E_{k}(N) >E_{WS}$ for $N\geq 
7 \> \, \mathrm{atoms}\cdot\mathrm{nm^{-3}}.$ 

This effective cross section can be compared with the effective 
one $\langle \sigma_{L}\rangle $ obtained in liquid Ar by Christophourou {\sl et al.} 
\cite{christo}. Even there,  $\langle \sigma_{L}\rangle $ appears to be 
much narrower and its minimum occurs at an energy 
much smaller than in the atomic one. An agreement with the two 
effective cross sections could be obtained 
by adding $E_{k}(N)$ to and 
by using $\mathcal{F}(k)$ rather than $S(0)$ in 
the data of Christophorou. 

\section{Conclusions}\label{fine}

In this paper measurements of the excess electron mobility in dense Ar 
gas in the neighborhood of the liquid--vapor critical point are
 presented. The most important result of the present experiment is the 
observation of a sharp maximum of the zero--field density--normalized 
mobility $\mu_{0}N$ at the same density where it occurs in the 
liquid. 

The interpretation of these measurements is challenging 
because two opposite points of views must be reconciled. In fact, in the 
low--density gas it is customary to adopt a single scattering picture, 
while in the dense liquid the electron transport properties are 
described in terms of scattering of electrons off collective 
excitations of the medium. The interesting point is to understand if 
the physical mechanisms underlying the scattering processes gradually change at 
some density between the dilute gas and the dense liquid or if the 
kinetic picture valid at low density can be still retained, with 
obvious modifications so as to include multiple scattering, also in 
the liquid phase.

To this goal, the present data have been analyzed by extending the 
heuristic model proposed by Borghesani {\sl et al.} to explain the 
mobility measurements in moderately dense Ar gas \cite{bsl}. The 
model is a density--modified kinetic model based on the classical 
kinetic theory \cite{hux}, where density--dependent multiple--scattering 
effects are included in a heuristic way. 
The model is based on the quantum density-dependent shift of the 
ground state energy of the electrons in a dense and disordered medium, 
on the accounting for correlations among scatterers described by the 
static structure factor of the medium, and on the quantum 
self--interference of the electron wavepacket scattered off randomly 
located scattering centers along paths connected by time--reversal. 
The kinetic term of the ground state energy shift must be added to 
the usual kinetic energy of the electrons when the cross section and 
other dynamic properties of scattering have to be evaluated.
All these effects concur to dress the atomic cross section resulting 
in an effective density--dependent momentum transfer cross section, 
that nonetheless is
intimately related to the atomic one. 

Although the data for quite high densities fit well in this model, the 
description of the newly discovered feature, namely the mobility 
maximum, requires the introduction of a density--dependent adjustable 
parameter that scales the cross section. In any case, however, the 
kinetic picture is preserved and the resulting effective cross 
section turns out to be very similar to the atomic one. In 
particular, it shows a Ramsauer--type minimum. 

The overall success of this kinetic description is striking, even more 
if one takes into account the limited accuracy with which the atomic 
cross section is known, especially in the region of the Ramsauer 
minimum, and the uncertainty with which the energy--dependent 
structure function $\mathcal{F}(\epsilon)$ is known, particularly in 
the neighborhood of the critical point. It also appears that 
more refined theoretical calculations of the kinetic energy shift at 
high density are needed, as well as a treatment of the effect of the 
density fluctuations on the electron energy distribution function.  

Moreover, these data still raise the question of how to treat
theoretically the scattering processes for momentum transfer at very 
large densities, possibly including contributions from mechanisms 
different from density--modified kinetic processes.

\newpage

\end{document}